\documentclass[prl,twocolumn,showpacs,preprintnumbers]{revtex4}
\usepackage{graphicx,amsmath}
\usepackage[dvipdfm,colorlinks=true, citecolor=blue, urlcolor=blue ]{hyperref}

\begin{document}

\title{Parity symmetry and parity breaking in the quantum Rabi model with addition of Ising interaction}
\author{Q. Wang$^{1,2}$}
\author{W. L. Yang$^{1}$}
\author{T. Liu$^{1,3}$}
\author{M. Feng$^{1}$}
\email{mangfeng@wipm.ac.cn}
\author{K. L. Wang$^{4}$}
\affiliation{$^{1}$ State Key Laboratory of Magnetic Resonance and Atomic and Molecular
Physics, Wuhan Institute of Physics and Mathematics, Chinese Academy of
Sciences, Wuhan 430071, China}
\affiliation{$^{2}$ College of Physics and Electronics, Hunan University of Arts and Science,
Changde 415000, China}
\affiliation{$^{3}$ The School of Science, Southwest University of Science and
Technology, Mianyang 621010, China}
\affiliation{$^{4}$ The Department of Modern Physics, University of Science and
Technology of China, Hefei 230026, China}

\begin{abstract}
We explore the possibility to generate new parity symmetry in the quantum Rabi
model after a bias is introduced. In contrast to a mathematical
treatment in a previous publication [J. Phys. A \textbf{46}, 265302
(2013)], we consider a physically realistic method by involving an
additional spin into the quantum Rabi model to couple with the original spin
by an Ising interaction. The rule can be found
that the parity symmetry is broken by introducing a bias and then restored by
adding new degrees of freedom. Experimental
feasibility of realizing the models under discussion is investigated.
\end{abstract}

\maketitle

{\it Introduction}.-  As one of milestones in the history of quantum
physics, the well-known quantum Rabi model (QRM) \cite {rabi} strictly describes
the simplest interaction between matter and the quantum light, which
has been widely employed to study great variety of physical systems,
such as trapped ions \cite{ion}, cavity and circuit quantum
electrodynamics \cite{cavity,circuit,cir2} as well as photonic
systems \cite{photon}.

Recently, much attention has been paid to the QRM for seeking the
closed-form analytical solution and the intrinsic characteristic
\cite {solu4,irish,Liu2,Braak,solu5,solu2,solu3}. Although the discrete $Z_{2}$
symmetry in the QRM makes the excitation number no longer as a
conserved quantity, we are still able to take the QRM as an
integrable system by considering the parity conservation \cite{Braak}. Due to
this fact, a parity chain in the QRM has been found in two
infinite-dimensional Hilbert invariant subspaces \cite{pchain},
which could be further extended to the $N$-state case \cite{Ncase}.

However, the situation turns to be completely different if a biased
field is introduced into the QRM, and we may call it as a biased
Rabi model (BRM) with the following form in units of $\hbar =1$,
\begin{equation}
H_{B}=-\Delta \sigma^{x}+\varepsilon \sigma^{z}+\omega a^{\dagger
}a+\lambda (a^{\dagger}+a)\sigma^{z},  \label{1}
\end{equation}
where $\Delta $ and $\varepsilon $ are the tunneling and the local
bias field, respectively, $\omega $ and $a^{\dagger }$ ($a$) are
frequency and the creation (annihilation) operator of the
single-mode bosonic field, and $ \lambda$ is the Rabi frequency.
$\sigma^{z,x}$ are the usual Pauli operators for the spin-1/2 and
$\sigma^{x}=\sigma^{+}+\sigma^{-}$ with $\sigma^{\pm
}=(\sigma^{x}\pm i\sigma^{y})/2$. Please note that the QRM can be
described by various Hamiltonians, e.g., unitarily transforming Eq.
(1) by $U=\frac{1}{\sqrt{2}}\left(\begin{array}{ll}
1 & 1  \\
-1 & 1
\end{array}\right)$ \cite{Braak}. But we work throughout this paper
by taking the quantization axis defined as in Eq. (1).

Compared to the standard form of the QRM, the additional bias term
$\varepsilon \sigma^{z}$ in Eq.(1) brings in complication but more
physics. For example, the parity symmetry in the QRM is broken due to the
addition of this bias \cite{Braak,Liu1}. If we define a parity
operator $P_{1}=\sigma ^{x}\otimes P_{0}$ with $P_{0}=e^{i\pi
a^{\dag}a}$, we may find $[P_{1}, H_{B}]\neq$0. Although this parity
breaking in the BRM can present us some interesting physics, such as
observation of unique scaling behavior and further understanding of the
rotating-wave approximation \cite{Liu1}, it is natural for us to
think of the possibility of restoring the broken symmetry. We have
noticed a very recent proposal \cite{GD} for a new nonlocal
symmetry in the BRM, implying a generalized parity. The main idea
is the introduction of a transformation $P$, enabling
$\varepsilon\rightarrow -\varepsilon $, $\sigma^{z}\rightarrow
-\sigma^{z}$ and $a(a^{\dagger})\rightarrow -a(-a^{\dagger})$.
Although it really commutes with $H_{B}$, $P$ is not a physically
meaningful operator to the BRM as described in \cite{GD} because it requires an additional
degree of freedom to be involved to carry out
$\varepsilon\rightarrow -\varepsilon$ and also lacks concrete models for justification.

In the present work, we focus on a physical consideration of a new symmetry in the BRM.
The key idea is the involvement of an additional spin coupled with the original spin by an Ising
interaction. This restoration of parity symmetry can be straightforwardly extended to more spins
once the new symmetry is also broken by an additional bias. We will discuss the experimental
feasibility of demonstrating the new parity symmetry and the parity breaking in the QRM plus Ising model.

{\it New parity symmetry and parity breaking}.- By introducing an auxiliary spin into the QRM, we have
\begin{equation}
H_{2}=-\Delta \sum_{i=1}^{2}\sigma_{i}^{x}+\omega a^{\dag}a+\lambda (a^{\dag}+a)\sigma _{1}^{z}+
\varepsilon \sigma _{1}^{z}\sigma _{2}^{z},
\end{equation}
where $\sigma_{2}^{x,z}$ are the Pauli operators for the new spin
coupled to the original spin by Ising coupling. In this case,
$\varepsilon$ is the Ising coupling strength, rather than the bias strength.
For convenience of description in the following, we will mention the original spin as
the first spin in order to distinguish from the newly joined spins.
We may define a new parity operator
$P_{2}=\prod_{i=1}^{2}\sigma_{i}^{x}\otimes P_{0}$, which fulfills
$[H_{2},P_{2}]=0$. The key point for the physical feasibility of the
new symmetry lies in the fact that
$P^{\dagger}_{2}(\varepsilon\sigma_{2}^{z})P_{2}= -\varepsilon\sigma
_{2}^{z}$, rather than simply making $\varepsilon\rightarrow
-\varepsilon$ in \cite {GD}. In other words, the new parity for Eq.
(1) works only when a new degree of freedom is introduced.
\begin{figure}[htp] \center
\includegraphics[scale=0.5]{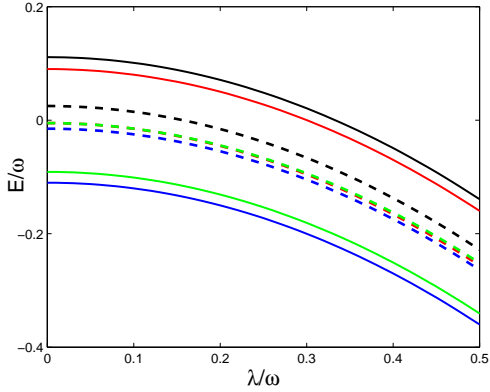}
\caption{(Color online) The four lowest eigen-energies of $H_{2}$ (dashed, $\eta=0$) and $H_{2B}$
(solid, $\eta\neq0$) as functions of the coupling strength $\lambda$ in the case of
$\Delta/\omega=0.01$, $\varepsilon/\omega=0.005$ and $\eta/\omega=0.1$.}
\end{figure}

The new symmetry will also break down if we introduce a new local bias on the
first spin, such as $\eta\sigma_{1}^{z}$. In such a case, $H_{2}$ turns to be
$H_{2B}$ with
\begin{equation}
H_{2B}=-\Delta \sum_{i=1}^{2}\sigma_{i}^{x}+\omega a^{\dag}a+\lambda (a^{\dag}+a)\sigma_{1}^{z}
+\eta\sigma_{1}^{z} +\varepsilon\sigma_{1}^{z}\sigma_{2}^{z},
\end{equation}
and it is evident that $[H_{2B}, P_{2}]\neq$0. As known in \cite
{Liu1}, scaling behavior would appear at the critical point of the
parity breaking. To see the scaling behavior in the QRM plus Ising
model, we may diagonalize Eq. (3) by displaced Fock states
$|n\rangle_{A}=\frac{e^{-q^{2}/2}}{\sqrt{n!}}(a^{\dagger}+q)^{n}e^{-qa^{\dagger}}|0\rangle$
and $|n\rangle_{B}=\frac{e^{-q^{2}/2}}{\sqrt{n!}}
(a^{\dagger}-q)^{n}e^{qa^{\dagger}}|0\rangle$ with the displacement
variable $q=\lambda/\omega$ \cite{Liu1,Liu2}. As a result, the
eigenfunction of $H_{2B}$ is given by
\begin{equation}
|\Psi\rangle=|\downarrow\downarrow\rangle|\Phi_{1}\rangle+|\downarrow\uparrow\rangle|\Phi_{2}\rangle+
|\uparrow\downarrow\rangle|\Phi_{3}\rangle+|\uparrow\uparrow\rangle|\Phi_{4}\rangle,
\end{equation}
where
$\sigma^{z}|\uparrow\rangle(|\downarrow\rangle)=|\uparrow\rangle(-|\downarrow\rangle)$,
$|\Phi_{1}\rangle=\sum_{n}a_{n}|n\rangle_{B}$,
$|\Phi_{2}\rangle=\sum_{n}b_{n}|n\rangle_{B}$,
$|\Phi_{3}\rangle=\sum_{n}c_{n}|n\rangle_{A}$ and
$|\Phi_{4}\rangle=\sum_{n}d_{n}|n\rangle_{A}$, with the coefficients
$a_{n}$, $b_{n}$, $c_{n}$ and $d_{n}$ to be determined by later calculation. So we
have to solve following Schr{\"{o}}dinger equations
\begin{eqnarray}
\sum_{n}(-1)^{m}D_{mn}(\varepsilon d_{n}-\Delta c_{n})+[\omega(m-q^{2})-\eta]a_{m} \nonumber\\
-\Delta b_{m}=Ea_{m}, \nonumber\\
\sum_{n}(-1)^{m}D_{mn}(\varepsilon c_{n}-\Delta d_{n})+[\omega(m-q^{2})-\eta]b_{m}  \nonumber\\
-\Delta a_{m}=Eb_{m}, \nonumber\\
\sum_{n}(-1)^{n}D_{mn}(\varepsilon b_{n}-\Delta a_{n})+[\omega(m-q^{2})+\eta]c_{m}  \nonumber\\
-\Delta d_{m}=Ec_{m}, \nonumber\\
\sum_{n}(-1)^{n}D_{mn}(\varepsilon a_{n}-\Delta b_{n})+[\omega(m-q^{2})+\eta]d_{m}  \nonumber\\
-\Delta c_{m}=Ed_{m}, \nonumber
\end{eqnarray}
with $E$ the eigenenergy and $D_{m,n}$ defined as \cite{Liu1,Liu2}
$$D_{m,n}=e^{-2q^{2}}\sum_{k=0}^{\min [m,n]}(-1)^{-k}\frac{\sqrt{m!n!}(2q)^{m+n-2k}}
{(m-k)!(n-k)!k!}.$$

We may obtain analytical solutions of the eigenenergies from above equations
under the condition of $\Delta/\omega\ll$1 \cite {explain}, for which the diagonal
terms of $D_{m,n}$ play dominant roles with respect to the
off-diagonal terms. In such a case, the eigenenergies are given by
$E_{m1}^{\pm}=-\Delta\pm\sqrt{D_{mm}^{2}(\varepsilon-\Delta)^{2}+\eta^{2}}+\omega(m-q^{2})$,
and
$E_{m2}^{\pm}=\Delta\pm\sqrt{D_{mm}^{2}(\varepsilon+\Delta)^{2}+\eta^{2}}+\omega(m-q^{2})$,
with $m=0, 1, \cdots$, where $E_{01}^{-}$ is the ground-state
eigenenergy. Fig. 1 plots the lowest four eigenenergies, from which
we know that the introduction of the bias into H$_{2}$, breaking down the the
parity of the original system, shifts the eigenenergies in a
symmetric way. i.e., half of the eigenenergies being lower and half being
higher. As a result, the ground-state eigenenergy is lower after the bias
is introduced.

{\it Scaling behavior}.- Using the ground-state eigenfunction, we obtain,
\begin{eqnarray}
\langle\sigma_{1}^{z}\rangle=\frac{-\kappa}{\sqrt{\kappa^{2}+e^{-4\beta}}},
\end{eqnarray}
with $\beta=q^{2}$ and $\kappa=\eta/(\Delta-\varepsilon)$. Compared
with the relevant results in \cite{Liu1}, the Ising coupling
strength $\varepsilon$ is involved in $\kappa$, which would definitely
modify the scaling behavior. Following the steps in \cite {Liu1}, we
define a scale $\beta_{c}=-\ln(2\kappa^{2})/4$ and a displaced scale
$\alpha=(\beta-\beta_{c})/\sqrt{27}$, and then we have
\begin{equation}
\langle\sigma_{1}^{z}\rangle=\frac{-\kappa}{\sqrt{\kappa^{2}+(2\kappa^{2})^{\beta/\beta_{c}}}},
\end{equation}
and
\begin{equation}
\langle\sigma_{1}^{z}\rangle=-1/\sqrt{1+2e^{-12\sqrt{3}\alpha}},
\end{equation}
the latter of which is independent of $\kappa$ and shows scaling invariance.

For a fixed value of $\kappa$, $\langle\sigma_{1}^{z}\rangle$ in Eq.
(6) is only relevant to the variable $\beta$, rather than to other
characteristic parameters. So $\beta_{c}$ can be regarded
as a scale of the QRM. Different from in \cite {Liu1},
however, the added Ising coupling leads to a bifurcation in the
scaling behavior, as shown in Fig. 2(a) where the lower (upper)
branch corresponds to $\Delta>\varepsilon$ ($\Delta <\varepsilon$).
In addition, $\beta=\beta_{c}$ corresponds to fixed crossing points
with the variation of $\beta$, in which $\langle\sigma_{1}^{z}\rangle$
turns out to be constants $\pm 1/\sqrt{3}$, i.e., the fixed crossing
points existing in the two branches. After a scaling
displacement, Eq. (7) is of the same form as in \cite{Liu1} and the
variation with $\alpha$ is formally independent of $\kappa$. As shown
in Fig. 2(b), the effect of the Ising coupling is reflected
in different values of $\alpha$ and $\langle\sigma_{1}^{z}\rangle$ in the curve.

\begin{figure}[htp] \center
\includegraphics[scale=0.5]{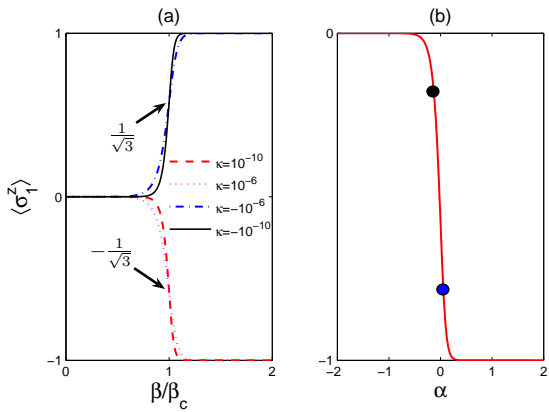}
\caption{(Color online) Scaling behavior of the ground-state $\langle\sigma_{1}^{z}\rangle$.
(a) As a function of $\beta/\beta_{c}$, which shows a bifurcation depending on the
difference between $\Delta$ and $\varepsilon$; (b) As a function of $\alpha$, which remains
unchanged with respect to different values of $\kappa$. The black (upper) dot means
$\eta/\Delta=10^{-6}$ and $\varepsilon=0$, and the blue (lower) dot represents $\eta/\Delta=10^{-6}$
and $\varepsilon/\Delta=0.5$ in the case of $q =$0.2.}
\end{figure}

{\it Discussion}.- Our treatment above can be generalized to the
N-spin case with one spin under QRM and coupled to other
$N-1$ spins by Ising interactions in a star configuration, which is
given by
\begin{eqnarray}
H_{N}=-\Delta\sum_{i=1}^{N}\sigma_{i}^{x}+\omega a^{\dag}a+\lambda(a^{\dag}+a)\sigma^{z}_{1}  \nonumber \\
~~~~~~~~~~~~~~~~~~~~~~~~~~~~~~+\sigma_{1}^{z}\otimes\sum_{k=2}^{N}\varepsilon_{k}\sigma_{k}^{z},
\end{eqnarray}
where $\varepsilon_{k}$ is the Ising coupling strength of the first
spin with the $k$th one. $H_{N}$ possesses a parity symmetry with
the corresponding parity operator
$P_{N}=\prod^{N}_{i=1}\sigma^{x}_{i}\otimes P_{0}$, due to
$[H_{N},P_{N}]=0$. As described above, this symmetry will be broken
by an additional local bias on the first spin, such as $\eta\sigma^{z}_{1}$, and
the parity breaking leads to scaling behavior of the ground state
similar to Eq. (5), but with
$\kappa=\eta/(\Delta-\sum_{k=2}^{N}\varepsilon_{k})$ in the present
case. Evidently, a new parity symmetry will appear once a new spin
moves in and turns the bias to be an Ising coupling to the first
spin. Simply speaking, it is a rule that the parity symmetry and
breaking appear alternately in the QRM by introducing a term of
Ising interaction and a term of bias.

\begin{figure}[htp] \center
\includegraphics[scale=0.5]{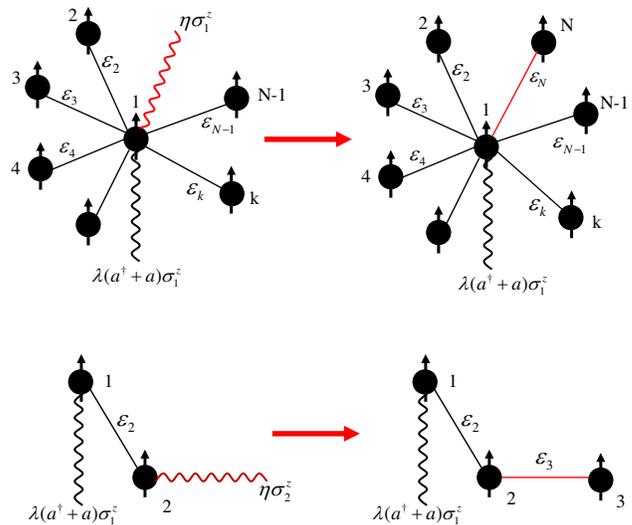}
\caption{(Color online) Upper panel: Sketch for star configuration, where the bias is applied
on the first spin to break down the parity symmetry and a new spin is introduced to generate
a new parity symmetry by turning the bias
to be an Ising interaction with the first spin; Lower panel: Sketch for linear configuration,
where the bias is applied on the newly joined spin (e.g., the second spin) to break down the
parity symmetry and an additional spin is introduced to generate
a new parity symmetry by turning the bias
to be an Ising interaction with the second spin.}
\end{figure}

However, what we described above is for a multi-spin Ising coupling
in a star configuration (See Fig. 3(a)), where the newly introduced spins
only couple to the first spin, and no coupling between any two of the
newly joined spins is assumed. What about other configurations,
such as the linear structure in Fig. 3(b) with the bias applied on the
newly joined spin? We find by straightforward deduction that the rule above still
works in this case, and the scaling behavior relevant to the parity breaking is also observable
if we measure $\langle\sigma_{1}^{z}\rangle$. The key point for observing the
scaling behavior is that our measurement should be made on the spin coupling directly to the
quantized field under the QRM. For other spins without direct
couplings to the QRM field, no scaling behavior can
be observed on them in the parity breaking.

Since both QRM and Ising model are usually employed interactions in
different fields of physics, we may achieve the models described
above and observe the predicted behavior with current laboratory
technique. Taking the ion-trap system as an example, we first
consider a single ultracold ion confined in the pseudo-potential of
a Paul trap under laser irradiation in a Raman $\Lambda$-type
configuration, whose hamiltonian in a frame rotating with the laser
frequency is given by \cite {solu4},
\begin{equation}
H_{ion}=\frac {\tilde{\Delta}}{2}\sigma^{z}+ \tilde{\nu} a^{\dagger}a + \frac
{\tilde{\Omega}}{2}[\sigma^{+}e^{i\tilde{\eta}(a^{\dagger}+a)} +
\sigma^{-}e^{-i\tilde{\eta}(a^{\dagger}+a)}],
\end{equation}
where $\tilde{\Delta}$ is the detuning of the laser to the two levels of the
spin, $\tilde{\nu}$ is the trap frequency with $a^{\dagger} (a)$ the
creation (annihilation) operator of the vibrational mode and
$\sigma^{z,\pm}$ are the usual Pauli operators for the spin.
$\tilde{\Omega}$ is the Rabi frequency and $\tilde{\eta}$ is the Lamb-Dicke
parameter. As shown in \cite {solu4,Liu1}, $H_{ion}$ can turn into a
similar form to Eq. (1) after some unitary transformation, where the bias
is relevant to the detuning $\tilde{\Delta}$. To achieve the Ising model, we
introduce another ion coupling to the first ion as,
\begin{equation}
H_{cc}=\omega_{s} S^{z} + \tilde{\varepsilon}\sigma^{z}S^{x},
\end{equation}
where $S^{z,x}$ are the usual Pauli operators for the new spin,
$\omega_{s}$ is the splitting frequency of the new spin with
coupling strength $\tilde{\epsilon}$ to the first one. The coupling
$\sigma^{z}S^{x}$ and similar forms of Ising coupling have been achieved
experimentally by off-resonant lasers and resonant Raman beams in trapped-ion systems \cite {kim1, kim2, islam, blatt2}.
Such couplings can also be generated by a magnetic field gradient \cite {wunder} or by a
non-uniform laser field \cite {porras1} on the trapped ions.
Following the unitary transformations in \cite{solu4} and
meanwhile performing a Hadamard gate on the new spin for
$S^{z}\Leftrightarrow S^{x}$, we may reach
\begin{eqnarray}
H'_{ion}= -\frac {\tilde{\Omega}}{2}\sigma^{x}+ \omega_{s}S^{x}+ \tilde{\nu}
a^{\dagger}a + \frac{\tilde{\nu}\tilde{\eta}}{2}(a^{\dagger}+a)\sigma^{z} - \frac {\tilde{\Delta}}{2}\sigma^{z} \nonumber
\\  +\tilde{\varepsilon}\sigma^{z}S^{z}, \nonumber
\end{eqnarray}
which is of the same form as in Eq. (3).  So the parity symmetry and the parity
breaking can be achieved experimentally by tuning $\tilde{\Delta}=0$ and $\tilde{\Delta}\neq$0, respectively.

The models under consideration are also feasible in circuit QED systems \cite {qed1,qed2,qed3} and
optomechanical system \cite {opto}, as exemplified in \cite {Liu1}, by introducing an
auxiliary spin coupling to the first spin by Ising interaction. A previous
publication has shown a quantum nondemolition detection by an auxiliary spin through
such an Ising coupling for light-matter interaction in a superconducting system \cite {diniz}.

{\it Conclusion}.- We have investigated the parity symmetry and parity
breaking relevant to the QRM by considering involvement of new spins. 
We obtained a general rule and also explored the scaling
behavior occurring in the case of the parity breaking, which strongly
depends on the Ising interaction. The
experimental feasibility to demonstrate the models and the unique
behavior is discussed. We believe that our results would be helpful
for further understanding light-matter interaction.

This work is supported by NFRPC (Grant No. 2012CB922102)
and NNSFC (Grants No. 11274352, No. 11274351 and No. 11347142).


\begin{thebibliography}{99}
\bibitem{explain} The reason that we consider the solution of the Schr{\"o}dinger equation in the
case of $|\Delta|/\omega \ll$1 is to obtain analytical expressions of the eigenenergy and
eigenfunction of Eq. (3), which helps writing the analytical expressions Eqs. (5)-(7) and helps
understanding the scaling behavior. This condition
is also acceptable experimentally, as discussed later.
\end{thebibliography}
\end{document}